\documentclass[conference]{IEEEtran}
\IEEEoverridecommandlockouts
\usepackage{comment}
\usepackage{cite}
\usepackage{amsmath,amssymb,amsfonts}
\usepackage{algorithmic}
\usepackage{graphicx}
\usepackage{capt-of} 
\usepackage{textcomp}
\usepackage{xcolor}
\usepackage{censor}
\def\BibTeX{{\rm B\kern-.05em{\sc i\kern-.025em b}\kern-.08em
    T\kern-.1667em\lower.7ex\hbox{E}\kern-.125emX}}

\makeatletter
\newcommand{\linebreakand}{%
  \end{@IEEEauthorhalign}
  \hfill\mbox{}\par
  \mbox{}\hfill\begin{@IEEEauthorhalign}
}
\makeatother

\begin{document}
\newcommand{\systemname}{CrowdForge}

%\title{AI and Mixed Reality-Mediated Audience Participation in Musical Performance\\
% \title{Audience Agency in AI- and Mixed Reality-Mediated Musical Performance\\
%\title{Evaluating Perceptual Impact of Audiovisual Elements in an AI- and Mixed Reality-Mediated Audience Participatory Performance\\
\title{\systemname{}: Effects of Audio and Visual Audience Participation Modalities on Agency, Collaboration, and Engagement in AI-Mediated MR Performances
\thanks{This work has been supported by the MUSMET project funded by the EIC Pathfinder Open scheme of the European Commission (grant agreement n. 101184379).}
}

% \author{  \IEEEauthorblockN{Anonymous Authors}
%     \IEEEauthorblockA{Anonymous Institutions}
\author{
\IEEEauthorblockN{Torin Hopkins}
\IEEEauthorblockA{\textit{SolJAMM Research and ATLAS Institute}\\
\textit{University of Colorado, Boulder}\\
Boulder, USA \\
TorinHopkins@gmail.com\\
}
%0000-0001-7359-2906}
\and
\IEEEauthorblockN{Shih-Yu Ma}
\IEEEauthorblockA{\textit{University of Colorado, Boulder} \\
Boulder, USA \\
Shih-Yu.Ma@colorado.edu\\
0000-0001-6686-0079}
\and
\IEEEauthorblockN{Zander Trafton-Hancock}
\IEEEauthorblockA{\textit{ATLAS Institute} \\
\textit{University of Colorado, Boulder}\\
Boulder, USA \\
zatr2033@colorado.edu}
\linebreakand
\IEEEauthorblockN{Ellen Yi-Luen Do}
\IEEEauthorblockA{\textit{ATLAS Institute} \\
\textit{University of Colorado, Boulder}\\
Boulder, USA \\
ellen.do@colorado.edu}
\and
\IEEEauthorblockN{Luca Turchet}
\IEEEauthorblockA{\textit{Dep. of Information Engineering } \\
\textit{and Computer Science}\\
\textit{University of Trento}\\
Trento, Italy \\
luca.turchet@unitn.it}
}

% \makeatletter
% \AtBeginDocument{
%   \let\@oldmaketitle\@maketitle
%   \renewcommand{\@maketitle}{
%     \@oldmaketitle
%     \vspace{-0pt} % Adjust spacing as needed
%     \begin{center}
%       \includegraphics[width=\linewidth]{IEEE_IS2_2025_IWMM_DAW_in_MR/figures/Performance_space.png}
%       \captionof{figure}{AIMAP performance with 3 performers and mixed reality projections.}
%       \label{fig:teaser}
%     \end{center}
%     \vspace{-5pt}
%   }
% }
% \makeatother

\maketitle

% \begin{figure*}[!t]
%     \centering
%     \includegraphics[width=\textwidth]{System_MRDAW.png}
%     \caption{Overview of the MRDAW mixed reality music collaboration system. (A) User n. 1 playing violin with a foot pedal for recording. (B) Mixed reality view from user showing the spatialized UI panel with track controls overlaid on the physical environment. (C) MR view from user demonstrating track recording feedback (red) and the spatial UI integrated into the physical space. (D) User n. 2 playing piano with a foot pedal for recording. Both users wear Meta Quest 3 headsets.}
%     \label{fig:system_overview}
% \end{figure*}

\IEEEpeerreviewmaketitle

\begin{abstract}

Technology-mediated audience participation (TMAP) enables audiences to influence live performances, yet balancing interaction with meaningful agency remains challenging. While AI-generated music and MR enable new forms of audio and visual audience participation, little is known about how these participation modalities influence audience experiences during live musical performance. In particular, it remains unclear how they shape perceived agency, collaboration, and engagement, and how they complement one another within TMAP systems. We present \systemname{}, an AI- and MR-mediated paradigm where audiences collectively control generative music and virtual environments during a live performance featuring a guitarist and DJ. Through a mixed-methods concert study evaluating three interaction conditions (audio, visual, and combined audio-visual), we demonstrate that observability is the primary driver of audience agency. While visual interaction provided easily observable feedback, audio parameters offered macro-structural control over the music. Audiences tended to use the control enabled by AI audio generation to test musical boundaries, introducing a productive adversarial tension between the crowd and the musicians. This dynamic pushed creative limits, deepened engagement, and generated moments of musical coherence when performers successfully resolved audience-induced challenges. We argue future TMAP systems should prioritize human-centered, highly observable interactions to build agency, while combining the expressiveness of music with the clarity of visual feedback.
\end{abstract}


%%%%%%%%%%%%%%%%%%%%%%%%%%%%%%%%%%%%%%%%%%%%%%%%%%%%%%%%%%%%%%%%%%%
\section{Introduction}
%\todo[inline]{add references where necessary}
% \textbf{from IRB}
% \todo[inline]{background history, main aims, and research questions FINISH and review}

\begin{figure*}
    \centering
    \includegraphics[width=\linewidth]{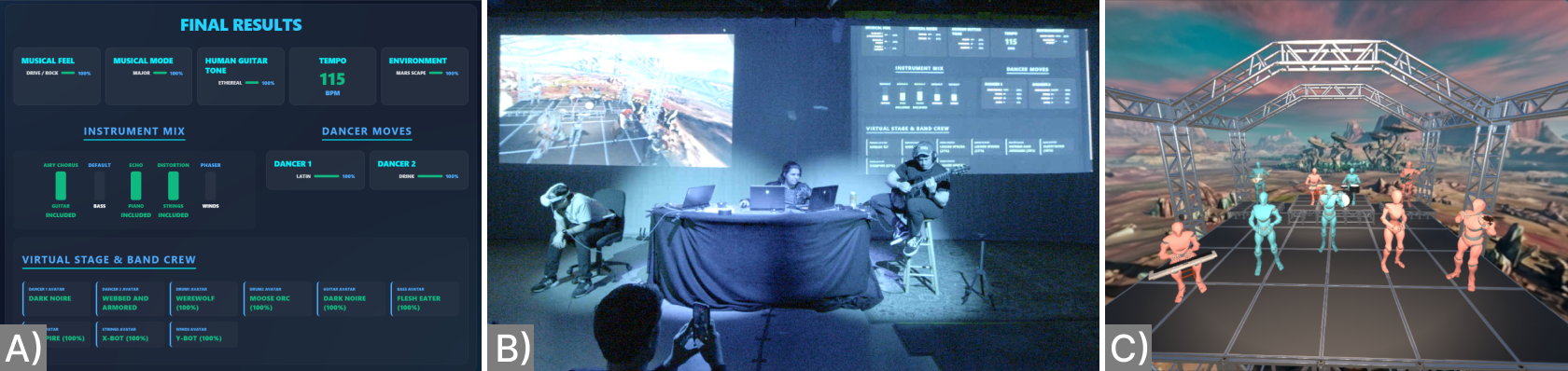}
    \caption{Overview of the audience participation workflow in \systemname. (A) Audience selections are aggregated into a collective configuration spanning musical character, tempo, instrumentation, dancer movement, virtual performers, and stage environment. (B) The selected parameters are incorporated into a live musical performance involving human musicians and projected audiovisual content. (C) The visual selections simultaneously configure a mixed-reality performance environment populated by virtual musicians and dancers. Together, these audio and visual modalities allow audience input to become a perceivable component of the resulting performance.}
    \label{fig:system_overview}
\end{figure*}

Traditional live music performances are predominantly unidirectional experiences, with minimal interaction between the audience and performers \cite{hopkins2024computer_dissertation}. Participatory music experiences shift this focus and prioritize aspects such as collaborative creativity and community building \cite{wu2017open, zhang2016web}. Though participatory music experiences are ubiquitous across cultures, and come in many different forms \cite{hopkins2024computer_dissertation, weinberg2003interconnected,amarie2025emoloop}, in recent years, the role of technology in facilitating audience participation has become an area of academic and artistic interest \cite{hodl2017TMAP, wu2017open, baldo2026sota_tmaps}. These interdisciplinary artistic works ask us to reconceptualize the traditional paradigm, shifting the experience of music performance towards an exploratory and communal effort. The field of technology-mediated audience participation (TMAP) in live music performance \cite{hodl2025tmap} simultaneously celebrates the artistry of trained live musicians and the use of technology as a means of communicating the intentions and sentiments of an audience, thus contributing to the creative process.

% \todo[inline]{cite projects throughout that correspond to these aspects}
Mozart famously utilized musical dice to promote audience interaction, having spectators roll dice to rearrange parts of the composition. As technologies became more sophisticated and interconnected \cite{hopkins2021user,turchet2018IoMusT}, engineers, designers, and artists alike have contributed to this growing field by utilizing a myriad of technological tools \cite{hodl2017TMAP,hodl2012designing, hodl2020large,jung2023hack,xambo2022mobile}. With the growing prevalence of networking, reliance on visual media (e.g., mixed reality, interactive visualizations), and complex engineered systems, TMAPs have begun to touch on many questions central to the musical metaverse \cite{turchet2023musical,rinaldi2024musical}.   

Recently, in \cite{hodl2025tmap} Hodl and Turchet outline several dimensions of technologies that circumscribe TMAPs, including light \cite{hopkins2023stringesthesia, freeman2005large}, physical devices \cite{Hopkins2023RhythmicConvos}, sensor-based devices \cite{hopkins2023stringesthesia, amarie2025emoloop,hirabayashi2015sense}, and smartphone/tablets \cite{zhang2016web, xambo2022mobile,weitzner2012massmobile}. This effort illustrated a comprehensive design space that connects the choice for various technological implementations, as well as the motivations for design decisions given the various technologies and aims of performance.  

From a complementary perspective, Wu et al. \cite{wu2017open} identified four key dimensions of TMAPs: agency (the perceived ability to influence the performance), participation modality (how audiences interact), degree of networking (the role of networked technologies in supporting participation), and scalability (the capacity to accommodate larger audiences).
%Coupled with this work, Wu et al. provided an exploration which classifies themes present in TMAPs, focusing largely on agency (i.e., the sense....), participation modality (i.e., the ....), degree of networking (i.e., the ....), and scalability (i.e., the ....) \cite{wu2017open}. 
Based on feedback from both audience members and performers, 
agency, participation modality, and interface design emerged as the most salient themes. Conversly, despite their presence in the TMAP framework proposed in \cite{hodl2025tmap}, little is known about how participation modality (e.g., interactive audio and visual elements) impacts the sense of agency, collaboration, and engagement. 

Additionally, many TMAP systems utilize visualization, sampling, and data points to impart change on musical and visual parameters, either providing information to performers (e.g., Open Symphony \cite{wu2017open}), or to inform musical content through sampling (e.g., Pharosphones \cite{ma2024pharosphones}). Some have even leveraged the variability in output potential of AI-generated music \cite{Hopkins2023RhythmicConvos, brade2026agentsinperformance} and Mixed Reality (MR) \cite{berthaut2015reflets_MR_TMAP}. However, to the best of our knowledge, there has yet to be a systematic presentation of how various creative and interactive modalities impact both audio/visual and perceptual outcomes.

Furthermore, there is an opportunity to explore how different participation modalities enabled by AI-generated music \cite{Hopkins2023Virtual, hopkins7brain, kim2026design_livemusicagents, hoffman2010shimon, bretan2016survey, brade2026agentsinperformance, baldo_masu_2026audienceNIME} and MR \cite{hopkins2024xrjam, ARDC_hopkins} 
influence audience experiences during live performance. In particular, we seek to understand how audio and visual participation modalities shape perceived agency, engagement, and social collaboration, and how they complement one another in supporting meaningful audience participation. This addresses a well-known challenge in designing TMAP systems: balancing accessible interaction (i.e., interfaces that are easy to understand and use) with opportunities for audiences to make meaningful creative contributions \cite{hodl2012designing, wu2017open, bin2018risky}. AI-generated music and MR are particularly well suited to enable such complementary participation modalities, allowing audiences to influence both musical and visual aspects of a live performance through intuitive, high-level interactions.

In this paper we present \emph{\systemname}, an AI and MR-mediated audience participation paradigm for live musical performance, as well as an accompanying experiment. We specifically aim to answer two research questions: 
\begin{enumerate}
\item  How do different audio and visual participation modalities impact an audience's sense of agency, engagement, and collaboration in a live music performance? 
\item How can the design of TMAP systems optimize the trade-off between making collaborative elements accessible to a diverse audience while ensuring they can make meaningful creative contributions?
\end{enumerate}

To answer such questions, we conducted a multi-condition, live performance featuring a DJ, a guitarist, an MR demonstrator in VR, and 19 interactive audience members. Following the performance \footnote{A recording of the performance is available: {\textcolor{blue}{https://www.youtube.com/watch?v=nbGJMI\_i9CU}}}, a mixed-methods evaluation was conducted.

% To answer such questions, we conducted a multi-condition, live performance featuring a DJ, a guitarist, an MR demonstrator in VR, and 19 interactive audience members. Following the performance \footnote{A recording of the performance is available  \href{https://www.youtube.com/watch?v=nbGJMI_i9CU}{\textcolor{blue}{https://www.youtube.com/watch?v=nbGJMI\_i9CU}}}, a mixed-methods evaluation was conducted.
Our study makes the following contributions:

\begin{enumerate}
    \item \systemname{}: A networked, AI-mediated audience participation performance paradigm in MR;
    \item A mixed-methods exploration of agency, collaboration, and engagement with respect to interactive modality (generated audio and MR visuals);
    \item Insights on how to promote a sense of agency, collaboration, and engagement in AI and MR-based TMAPs for live music performance.

\end{enumerate}

% The addition of a foot pedal also provides musicians with the ability to interact with recorded material in a familiar manner that simultaneously frees the hands to play their instrument. Beyond these aspects, musicians are able to work collaboratively on a digital audio workstation and interact with a looping mechanism—two features that have yet to be explored. 

\vspace{0.5cm}

% \section{Design of Participatory Live Music Performances}
% \input{IEEE_IS2_2025_IWMM_DAW_in_MR/01b_Related_Works}

%%%%%%%%%%%%%%%%%%%%%%%%%%%%%%%%%%%%%%
\section{System Design}
% \todo[inline]{system description and technical implementation}

\begin{figure}
    \centering
    \includegraphics[width=\linewidth]{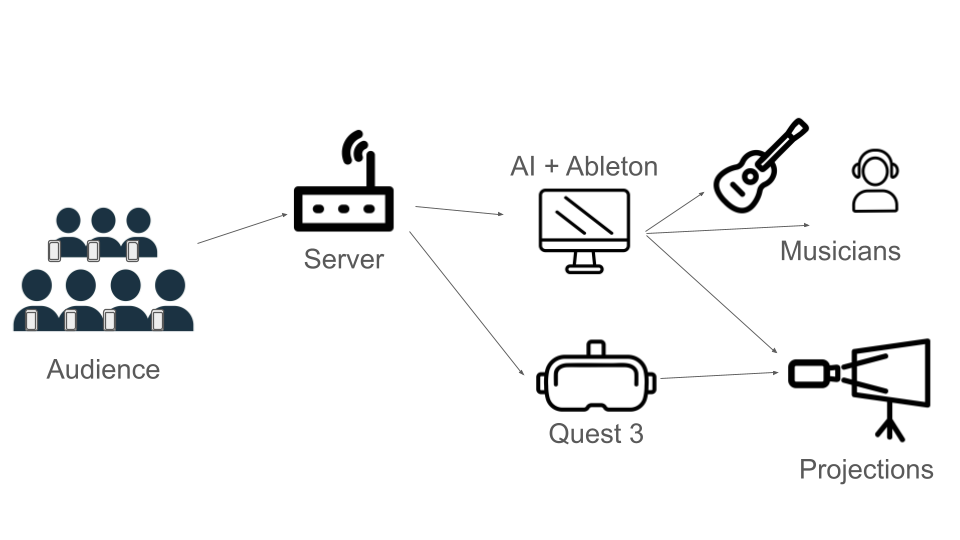}
    \caption{System diagram and signal flow illustrating the control audience members had over elements of the performance (from left to right). Audience members visit a web page on their mobile devices and vote on parameters they want to see reflected in the performance. The inputs are then aggregated and sent to both a generative transformer model (played through Ableton Live 12 software) and a mixed reality engine (Unity 3D) for rendering both audio and visuals.}
    \label{fig:system_setup}
\end{figure}

The \systemname{} performance system consisted of four main components: 1) the web interface for aggregating audience preferences, 2) AI model for music generation, 3) Ableton Live 12 for DJ and AI control, and 4) a projected MR environment (see Figures \ref{fig:system_overview} and \ref{fig:system_setup}).

In-person audience members connect to the server via their smartphones and vote on parameters of the performance at the beginning of each set (roughly every 7 minutes). Depending on the condition, participants voted on either audio parameters, visual parameters, or both. Audio parameters included aspects that impact the generated AI music (such as genre and tempo) and other audio parameters (e.g., guitar tone of a live human guitar player). Visual parameters impacted the projected MR visual scene, such as virtual characters and the virtual environment (see Table \ref{tab:conditions_separated}). 
On stage a guitar player, a DJ, and an MR demonstrator wearing a Quest 3 headset were featured. Though future performances can include audience members wearing headsets or other MR devices, the test performance only included an MR demonstration of a single perspective with a projected visual scene (see Figures \ref{fig:system_overview} and \ref{fig:system_setup}).    

% \begin{figure}
%     \centering
%     \includegraphics[width=\linewidth]{IEEE_IS2_2025_IWMM_DAW_in_MR/figures/AIMAP_system_diagram_simplified.png}
%     \caption{System diagram and signal flow illustrating the control audience members had over elements of the performance (from left to right). Audience members visit a web page on their mobile devices and vote on parameters they want to see reflected in the performance. The inputs are then aggregated and sent to both a generative transformer model (played through Ableton Live 12 software) and a mixed reality engine (Unity 3D) for rendering both audio and visuals.}
%     \label{fig:system_setup}
% \end{figure}

\subsection{Web Interface Architecture and User Interaction}
The web interface serves as the primary interactive conduit between the live audience and the generative AI performance system.
% The backend architecture is constructed on a Node.js runtime utilizing the Express.js framework for efficient HTTP routing and session management. To guarantee immediate, bidirectional state synchronization across all connected mobile clients, the system implements a WebSocket integration via Socket.io, facilitating synchronized navigation events.
From the audience perspective, users enter the ecosystem through an onboarding portal where they manipulate sliders and dropdown menus to cast votes on musical, visual, or both musical and visual parameters (depending on condition). 

% macro-musical parameters---such as genre, temporal characteristics, and instrument effects---as well as visual aesthetics, including 3D environments and avatar choreography depending on condition. As these inputs are submitted, the backend logs the telemetry for subsequent empirical analysis.

\begin{figure}
    \centering
    \includegraphics[width=\linewidth]{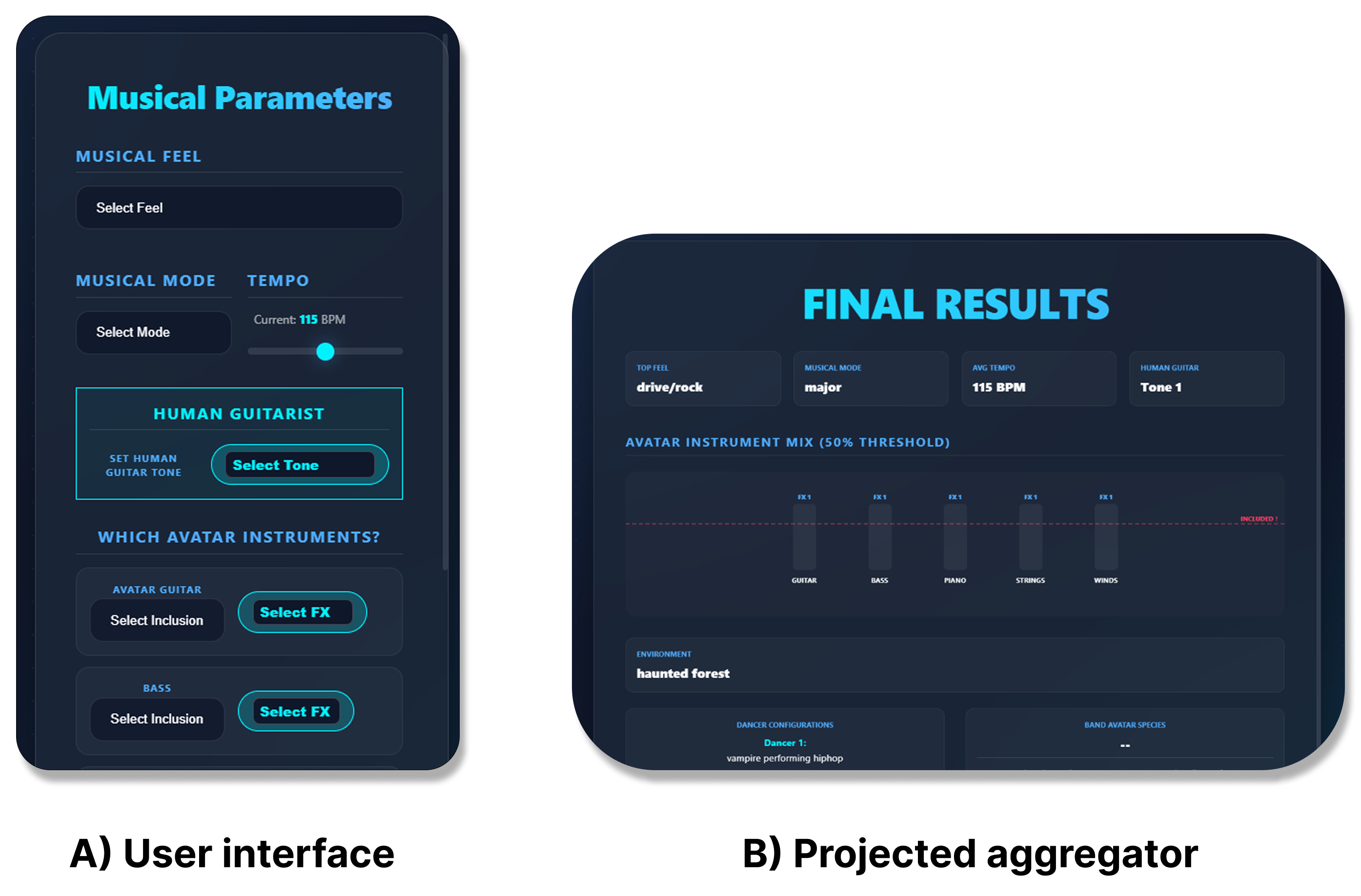}
    \caption{A) Mobile user interface for audience member during the audio parameter selection condition. B) Sample projected aggregator screen displayed to audience after selections were made.}
    \label{fig:UI}
\end{figure}

% To translate the audience's inputs into a cohesive performance, the server executes a real-time aggregation algorithm that determines the majority consensus for each variable. 
The finalized data structure is then visualized on a public-facing aggregator dashboard. The server then simultaneously broadcasts to the Ableton Live API for set control (via AbletonOSC \cite{jones2023abletonosc}), the causal model for multitrack MIDI generation, and the Unity Engine for 3D visual rendering (see Figures \ref{fig:UI} and \ref{fig:system_setup}).

% The architecture incorporates a fallback protocol to guarantee system stability; in instances of missing data or tied voting distributions, the backend algorithm utilizes randomization to autonomously complete the parameter set. 

\subsection{AI Model for Music Generation}

% \subsubsection{Base Architecture and Domain Adaptation}
% The system employs Llama-Z, a domain-adapted iteration of the Llama-3.2-1B causal language model. While the base Llama-3.2 contains 1.23 billion parameters, the model was modified via vocabulary expansion, elevating the total parameter count to 1.77 billion. To mitigate forgetting and ensure computationally efficient domain transfer, the training utilizes Low-Rank Adaptation. This configuration yields 539 million trainable parameters, representing 30.3\% of the expanded model. 

% The LoRA hyperparameters are strictly set to rank $r=16$ and scaling factor $\alpha=32$, applied alongside fully unfrozen embedding layers and the output head.
% Training execution utilizes BFloat16 precision for memory efficiency.

\subsubsection{AI Model and Tokenization Scheme}
The system employs Llama-Z, a domain-adapted iteration of the Llama-3.2-1B causal language model. The REMI-z tokenization scheme \cite{ou2026unifying} was utilized for its ability to generate multiple cohesive instrument parts in a fully orchestrated track while enabling audience members to omit an instrument and retain coherence in the generated output. Data representation relies on a specialized variation of the REMI-z tokenization protocol, namely the inclusion of genre. 

% This custom vocabulary contains 650 specific tokens mapping discrete musical events into a one-dimensional sequence comprehensible to the transformer. The vocabulary structure encapsulates:
% \begin{itemize}
%     \item \texttt{b-1}: Bar boundary markers.
%     \item \texttt{g-X}: Genre conditioning tokens representing four macro-stylistic clusters (e.g., \texttt{Z\_GROOVE}, \texttt{Z\_DRIVE}, \texttt{Z\_PULSE}, \texttt{Z\_ELEGANCE}).
%     \item \texttt{i-X}: Program/instrument declarations.
%     \item \texttt{o-X}: Temporal onset indicators mapped to a high-resolution grid of 128 positions per bar.
%     \item \texttt{p-X} and \texttt{d-X}: Pitch values and rhythmic durations, respectively.
% \end{itemize}

\subsubsection{Pre-processing and Mathematical Normalization}
To reduce the complexity of the data distribution and adhere to the REMI-z tokenization scheme, all incoming multitrack corpora---comprising 20,000 sampled segments from the Los Angeles MIDI Dataset (LAMD) and the Slakh2100 dataset---are normalized by discarding any segments not adhering to a $4/4$ time signature.

All harmonic content is mathematically shifted to a C-Major or A-Minor ground truth. Key detection is performed using the Krumhansl-Schmuckler (K-S) correlation algorithm, comparing an extracted 12-tone pitch class distribution histogram.

% , $\mathbf{h}$, against established perceptual profiles for major ($\mathbf{M}$) and minor ($\mathbf{m}$) keys. 

% The K-S profiles are defined as:
% \begin{align*}
%     \mathbf{M} &= [6.35, 2.23, 3.48, 2.33, 4.38, 4.09, 2.52, 5.19, 2.39, 3.66, 2.29, 2.88] \\
%     \mathbf{m} &= [6.33, 2.68, 3.52, 5.38, 2.60, 3.53, 2.54, 4.75, 3.98, 2.69, 3.34, 3.17]
% \end{align*}

% The optimal key index $k \in \{0, 1, \dots, 11\}$ and tonality ($major \in \{0, 1\}$) are determined by maximizing the Pearson correlation coefficient $r$:
% \begin{equation}
%     k_{\text{opt}}, major_{\text{opt}} = \arg\max_{k, major} \left( \frac{\text{cov}(\mathbf{h}, \text{roll}(\mathbf{P}_{major}, k))}{\sigma_{\mathbf{h}} \sigma_{\mathbf{P}}} \right)
% \end{equation}
% where $\mathbf{P}_1 = \mathbf{M}$, $\mathbf{P}_0 = \mathbf{m}$, and $\text{roll}(\dots, k)$ is a circular shift operator.

% Once the key is identified, a transposition integer $\Delta \in [-6, 6]$ is computed. For major keys, $\Delta = -k$. For minor keys, $\Delta = 9 - k$. Pitch events are subsequently shifted by $\Delta$ to ensure absolute harmonic normalization prior to tokenization.

\subsection{System Integration Pipeline: \\ Web Interface $\rightarrow$ AI $\rightarrow$ Ableton Live}
The inference architecture utilizes a multi-threaded pipeline translating web-based audience requests into arranged multitrack compositions inside Ableton Live. 

\subsubsection{Autoregressive Generation}
The generation sequence initiates when the web interface transmits an OSC message to the model to initiate generation given the input selected by the audience. The message specifically contains the target genre and the instrumentation (e.g., Drums, Piano, Guitar).

\subsubsection{The Ableton Controller Integration}
% Once the MIDI file is rendered, the generator passes the file directory via OSC. 

The controller script acts as a stateful agent directly manipulating Ableton Live's API, affecting parameters such as tempo and musical mode, as well as copying the content of locally generated MIDI files as launchable tracks in Ableton Live's session mode. This enables the DJ to manually manipulate the tracks as they are generated, affording control in a manner similar to pre-arranged tracks during live performance. 

% \begin{enumerate}
%     \item \textbf{Deterministic Channel Routing:} The controller parses the generated MIDI file, indexing active notes. It determines track routing using a hierarchical fallback map, routing Drums (Prog 128) to Track Base 1, Piano (Prog 0) to Track Base 7, etc.
%     \item \textbf{Dynamic Slot Manipulation:} To ensure no user data is overwritten, the controller communicates with Ableton. It iterates through scene rows until an entirely clear horizontal block (a ``slot'') is discovered.
%     \item \textbf{Clip Population and Synchronization:} The controller dynamically gauges the maximum duration of the generated notes, rounds to the nearest 4-bar multiple (capping at 16 bars), and issues commands to Ableton. Notes are injected directly into the DAW.
%     \item \textbf{Auto-Fire Resolution:} Once all clips are populated across their respective tracks, the controller executes a localized ``stop'' command  targeting only tracks associated with conflicting genre offsets. Finally, it sends synchronized \texttt{/live/clip/fire} commands, seamlessly transitioning the DAW's playback to the newly generated AI multitrack.
% \end{enumerate}

\subsection{Mixed Reality Viewer}

The MR viewer was developed in Unity 3D and deployed on Meta Quest 3 headsets as the primary visual-rendering component of the \systemname{} performance system. The viewer receives near-real-time control messages from the aggregation server and updates the visual scene according to the audience-selected parameters. These parameters include both character appearances and environmental themes, allowing the projected MR environment to reflect the collective preferences gathered through the web interface.

The MR scene contains a virtual ensemble consisting of two dancers, two drummers, one keyboardist, one violinist, one flutist, one guitarist, and one bassist (see Figure~\ref{fig:system_overview}C). Each instrumental character is configured to respond to OSC messages transmitted from the Ableton Live controller. These messages trigger synchronized performance animations, such as striking the appropriate drum pad, pressing keys within the correct keyboard region, or performing the corresponding instrumental gesture. The dancers are controlled through a combination of style commands from the aggregation server and timing messages derived from Ableton Live playback, enabling them to perform choreographic motions that remain synchronized with the music.

% Upon receiving the corresponding OSC message from the aggregation server, the Unity application updates the character models to match the selected visual configuration. In addition to these appearance changes, 

The environmental layer of the MR viewer places the full ensemble on a virtual stage. Similar to the character appearances, the active environment is updated through OSC messages from the aggregation server, allowing the visual scene to transition in response to audience voting results. 

% At the end of each performance session, the Unity scene resets to its default state to prepare for the next interaction cycle.

During the live performance, one Meta Quest 3 headset was used to project an example view point in VR. The demonstrators view point, gaze direction, and movements were cast to the projection display--enabling the audience to view the virtual space and the performance space simultaneously. Though viewing the VR environment could have been accomplished without a human demonstrator, the demonstrator provided the audience with a dynamic and human perspective within an immersive environment. A full view of the MR performance environment is shown in Figure \ref{fig:system_overview}.

% \section{AI Model Architecture \& Pre-processing Framework}
% \input{02a_AI_Model_Training}

\section{Study Design and Performance}

The study consisted of three music sets (roughly 7 minutes each) per performance differentiated by the selections the audience could make: 1) \textbf{audio only}, 2) \textbf{visual parameters only}, and 3) \textbf{both audio and visual parameters}. Audience members voted on performance parameters via web form before the beginning of each set. Two performances consisting of three conditions/sets occurred over the course of one day (N=19 audience members in total, N=5 for performance 1 and N=14 for performance 2). The first performance began with audio and the second with the visual condition. 

The study was approved by the University of Colorado Institutional Review Board and took place in a media studio, akin to a black box theater. The session began with a brief explanation of how to participate and all participants logged into the local area network to maintain connection throughout the performance. It was ensured that all participants were able to connect to the network before beginning the performance. Participants faced the performers and were instructed to make selections at the beginning of each condition/set. 

There were two live performers: 1) a guitar player, 2) a DJ, and an MR demonstrator wearing a Quest 3 headset and projecting their live point of view of the virtual stage to the audience. Behind the performers was the projection of the virtual environment and a projection of the aggregator screen (see Figure \ref{fig:system_overview}B). While the audience made selections (see Table \ref{tab:conditions_separated}) the aggregator screen showed user-chosen names for the audience contributors, enabling a dimension of self expression and a visual check to ensure all members of the audience remained connected to the network. 

After the audience members finished making their selections individually on their mobile devices the music generation began. This initiated the aggregation screen, which was immediately made visible to the audience, enabling them to see the results of the collective voting (see Figure \ref{fig:UI}). Within approximately 10 seconds the DJ began observing and playing clips which were generated (based on audience selections) and programatically populated in the Ableton Live set. The guitar player improvised with the DJ-selected generated tracks and each set lasted approximately 7 minutes. 

After the three sets concluded, the audience members were instructed to complete an online questionnaire. Following the questionnaire, six audience members total from the two performances volunteered to stay for a brief semi-structured audio-recorded interview (10-20 minutes). 

\begin{table*}[htbp]
\centering
\renewcommand{\arraystretch}{1.5}
\begin{tabular}{|p{0.35\textwidth}|p{0.6\textwidth}|}
\hline
\textbf{Condition (No. of Parameters)} & \textbf{Available Options} \\
\hline
\textbf{Audio (14)} \newline
- Genre / Feel \newline
- Tempo \newline
- Mode \newline
- Human Guitar Tone \newline
- Instruments \newline
- Piano FX \newline
- Guitar FX \newline
- Bass FX \newline
- Strings FX \newline
- Winds FX
&
\textbf{Genre / Feel:} drive/rock, pulse/hiphop, groove/funky, elegant/atmospheric \newline
\textbf{Tempo:} 90--140 BPM \newline
\textbf{Mode:} major, minor \newline
\textbf{Human Guitar Tone:} clean, distorted, atmospheric \newline
\textbf{Instruments:} guitar, bass, piano, strings, winds \newline
\textbf{Piano FX:} default, dist, echo \newline
\textbf{Guitar FX:} default, dist, airy \newline
\textbf{Bass FX:} default, dist, reverb \newline
\textbf{Strings FX:} default, dist, chorus \newline
\textbf{Winds FX:} default, dist, phaser \\
\hline
\textbf{Visual (14)} \newline
- Environment \newline
- Character Skins for 9 characters \newline
- Dance Moves for 2 dancers
&
\textbf{Environment:} haunted forest, mars scape, laser grid \newline
\textbf{Character Skins:} vampire, werewolf, moose orc, goblin elf, flesh eater, dark noire, lizard species, webbed and armored, x-bot, y-bot \newline
\textbf{Dance Moves:} hiphop, breakdance, latin, party \\
\hline
\textbf{Both (28)} \newline
- All Audio Parameters \newline
- All Visual Parameters
&
\textbf{Description:} The ``both'' mode combines all available parameters from the audio and visual conditions.  \\
\hline
\end{tabular}
\vspace{.01cm}
\caption{Parameters and Available Options per App Condition}
\label{tab:conditions_separated}
\end{table*}

% This condition gives users simultaneous control over the generative music system (tempo, genre, instrumentation, and effects) and the 3D rendering elements (environments, character skins, and animations).

%
%\begin{figure}[htbp]
%\centerline{\includegraphics{fig1.png}}
%\caption{Example of a figure caption.}
%\label{fig}
%\end{figure}
%

%%%%%%%%%%%%%%%%%%%%%%%%%%%%%%%%%%%%%%%%%%%%%%%%%%%%%%%%%%%%%%%%%%%
\section{Audience Member Evaluation}

Our evaluation is based on feedback from 19 total audience members ($N = 19$; 17 men, 2 women; $M_{age} = 28.89$, $SD = 9.89$ ) and the two musicians taking part in the performance. Audience member feedback was assessed through questionnaire (N=14, 5 opted out), behavior as recorded in the application (N=17, two experienced connection issues), and semi-structured interviews (N=6, of the 19 volunteered to stay for interviews). The performers provided self-reflective accounts of the performance experience. 

\subsection{Behavioral Analysis}
Skip rates provided behavioral validation of participants' engagement, aligning closely with the subjective ratings. When isolated in single-modality conditions, participants heavily favored the active domain. In the Audio condition, participants actively engaged with audio elements—showing a 0\% skip rate for tempo and a 6.7\% skip rate for genre and human guitar tone—while almost entirely ignoring the inactive visual configurations (93.3\% skip rate for characters, moves, and environments). In the Visual condition, participants engaged with visual settings (only an 11.1\% skip rate for characters and moves). In the ``Both" condition where all options were available, engagement balanced out but prioritized perceived high-impact features. Participants interacted heavily with core structural elements across both domains: tempo (0\% skip rate), visual characters and moves (7.7\% to 15.4\%), and genre/environment (15.4\%). More granular audio configurations—such as specific effects choices for AI instruments—saw much higher skip rates ranging from 30.8\% to 38.5\%. This behavioral data mirrors the subjective Influence and Agency findings, which suggest elements that participants felt gave them high influence and control (like tempo, genre, and human musician instrument tone) were rarely skipped, whereas the AI details that scored poorly on influence metrics were frequently bypassed even in a fully multimodal environment (see Figure \ref{fig:audioAgencyandInfluenceVneutral}).

% \begin{figure}[h]
%   \centering
%   \includegraphics[width=\linewidth]{IEEE_IS2_2025_IWMM_DAW_in_MR/figures/skipped_elements_both.png}
%   \caption{Graph indicating the top 10 most frequently skipped items in the supplied form during the condition when all parameters were made available to the audience. Y-axis demonstrates the item name, while the X-axis indicates the percentage of times the parameter was omitted.}
%   \label{fig:behavioral_skipped}
% \end{figure}

% \begin{figure}[h]
%   \centering
%   \includegraphics[width=\linewidth]{IEEE_IS2_2025_IWMM_DAW_in_MR/figures/behavioral_seconds_to_submit.png}
%   \caption{Bar graph indicating the number of seconds participants took to fill out the form. The audio and visual conditions both contain 14 parameters to choose from and the both condition contained 28.  There is no significant difference between the conditions.}
%   \label{fig:behavioral_seconds}
% \end{figure}

\subsection{Exploratory Questionnaire Results}

To understand how modality impacted the interactive audience experience, we analyzed participant responses ($N=14$) across four primary constructs: Sense of Agency, Social Presence, Emotional Response via the Self-Assessment Manikin (SAM), and the perceived influence of audio parameters.

To evaluate differences across the experimental conditions (Audio, Visual, Both) and elements, non-parametric statistical tests were employed due to the ordinal nature of the data and distribution assumptions. A Friedman Omnibus Test was first conducted to identify if there were overall significant differences among the related groups. When the Friedman test indicated significance (or near significance), Wilcoxon signed-rank tests were used for pairwise post-hoc comparisons. To account for the family-wise error rate, standard Bonferroni corrections were applied when evaluating the three main conditions (3 comparisons), while a step-down Holm-Bonferroni correction was applied when evaluating the audio sub-elements to preserve statistical power across the larger set of tests (15 pairwise comparisons among 6 elements).
Additionally, one-sample Wilcoxon signed-rank tests were used to evaluate whether absolute scores in each condition significantly deviated from a neutral baseline (e.g., 4 on a 1–7 Likert scale; 5 on a 1–9 SAM scale). Effect sizes for all Wilcoxon tests are reported as $r$, where values of $0.1$, $0.3$, and $0.5$ indicate small, medium, and large effects, respectively.

% To evaluate differences across the experimental conditions (Audio, Visual, Both) and elements, non-parametric statistical tests were employed due to the ordinal nature of the data and distribution assumptions. A Friedman Omnibus Test was first conducted to identify if there were overall significant differences among the related groups. When the Friedman test indicated significance (or near significance), Wilcoxon Signed-Rank Tests were used for pairwise post-hoc comparisons. Additionally, Wilcoxon Signed-Rank Tests were utilized to compare individual condition metrics against a neutral baseline. Effect sizes for the Wilcoxon tests are reported as $r$, where values of $0.1$, $0.3$, and $0.5$ indicate small, medium, and large effects, respectively. Where applicable, Bonferroni (for 3 condition comparisons) or Holm-Bonferroni (15 audio element comparisons) corrections were applied to account for multiple comparisons.
% Additionally, one-sample Wilcoxon signed-rank tests were used to evaluate whether absolute scores in each condition significantly deviated from a neutral baseline (4 on a 1–7 Likert scale; 5 on a 1–9 SAM scale). Effect sizes ($r$) are reported to indicate the magnitude of the observed effects.

\subsubsection{Sense of Agency}
To evaluate sense of agency, we analyzed responses from the Agency questionnaire alongside Dominance (i.e., perceived control) scores from the SAM. For the Agency measure, no significant overall differences were observed across the modalities (Friedman $p = 0.721$). 

% Pairwise comparisons confirmed that Agency scores did not significantly differ between the Audio and Visual ($p = 0.395, r = 0.23$), Audio and Both ($p = 0.329, r = 0.26$), or Visual and Both ($p = 0.271, r = 0.29$) conditions. This lack of relative difference was further contextualized when comparing these scores against a neutral baseline; neither the Audio ($p = 0.123$), Visual ($p = 0.975$), nor Both ($p = 0.073$) conditions yielded Agency scores significantly higher than neutral.

\begin{figure}[h]
  \centering
  \includegraphics[width=\linewidth]{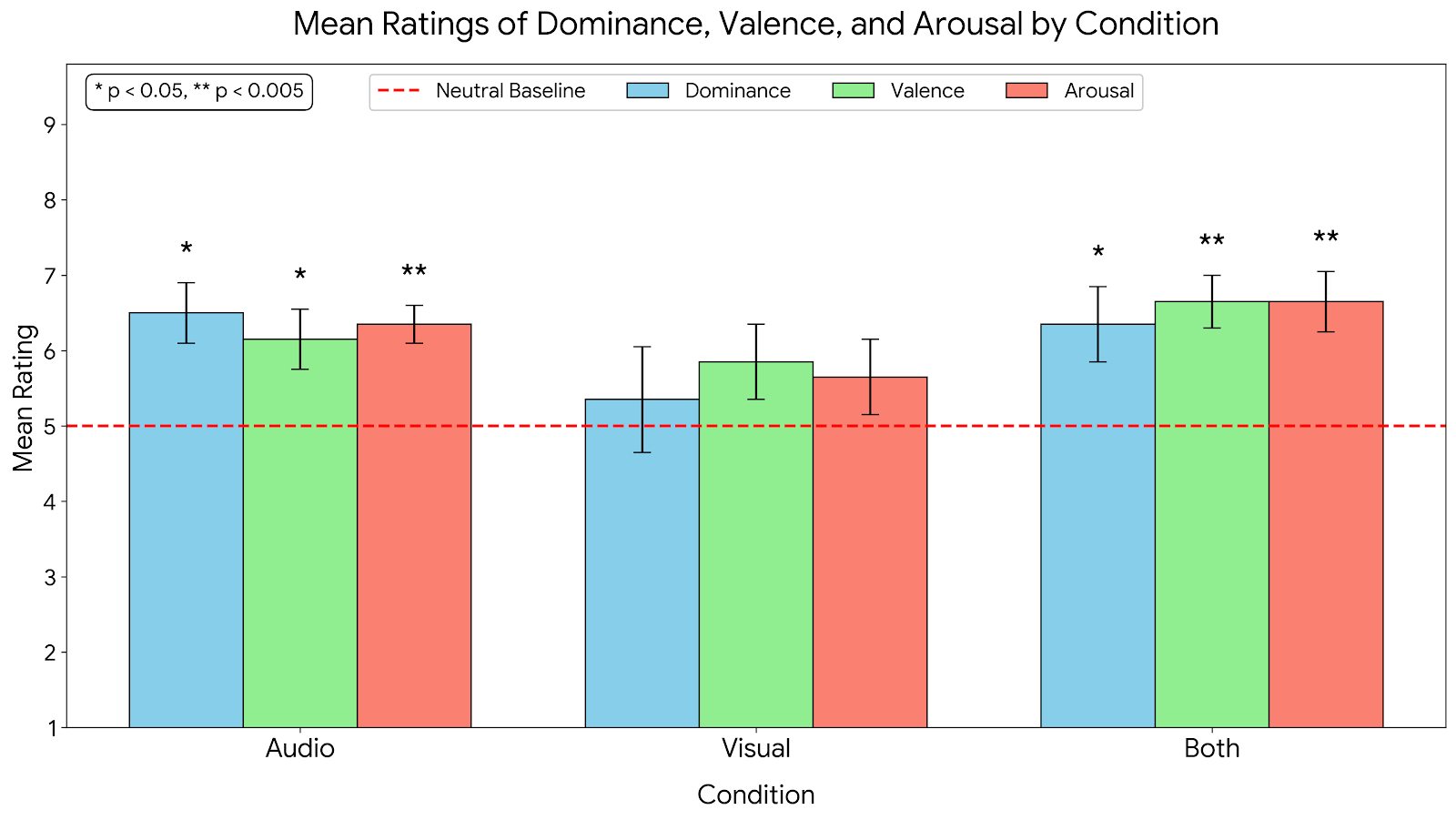}
  \caption{Indicates dominance, valence, and arousal scores on the self-assessment manikin, comparing Audio, Visual, and Both conditions. Each measure is rated 1-9. Significance figures are demonstrated with respect to the neutral baseline of 5 (shown in red).}
  \label{fig:dav_combined}
\end{figure}

However, the SAM Dominance scores—which measure the emotional feeling of being in control of the situation—demonstrated significant impact. While the omnibus test for Dominance did not show strict overall significance across conditions (Friedman $p = 0.368$), and pairwise differences were marginal (Audio vs. Visual $p = 0.075, r = 0.48$), comparisons against the neutral baseline revealed the Audio condition ($p = 0.006, r = 0.73$) and the Both condition ($p = 0.025, r = 0.60$) significantly elevated feelings of Dominance above the neutral baseline with large effect sizes (see Figure \ref{fig:dav_combined}). In contrast, the Visual-only condition did not elevate Dominance beyond neutral ($p = 0.612$). This indicates that while explicit, cognitive ratings of agency remained stable across modalities, the underlying affective sense of dominance and control was significantly driven by the presence of audio.

% \begin{figure}[h]
%   \centering
%   \includegraphics[width=\linewidth]{IEEE_IS2_2025_IWMM_DAW_in_MR/figures/dominance_vs_neutral.png}
%   \caption{Indicates dominance scores on the self-assessment manikin. Dominance is rated 1-9 and acts as a proxy for perceived control. Significance figures are demonstrated with respect to the neutral baseline of 5 (shown in red).}
%   \label{fig:domVneutral}
% \end{figure}

\subsubsection{Sense of Social Presence Between Audience Members}
\begin{figure}[h]
  \centering
  \includegraphics[width=\linewidth]{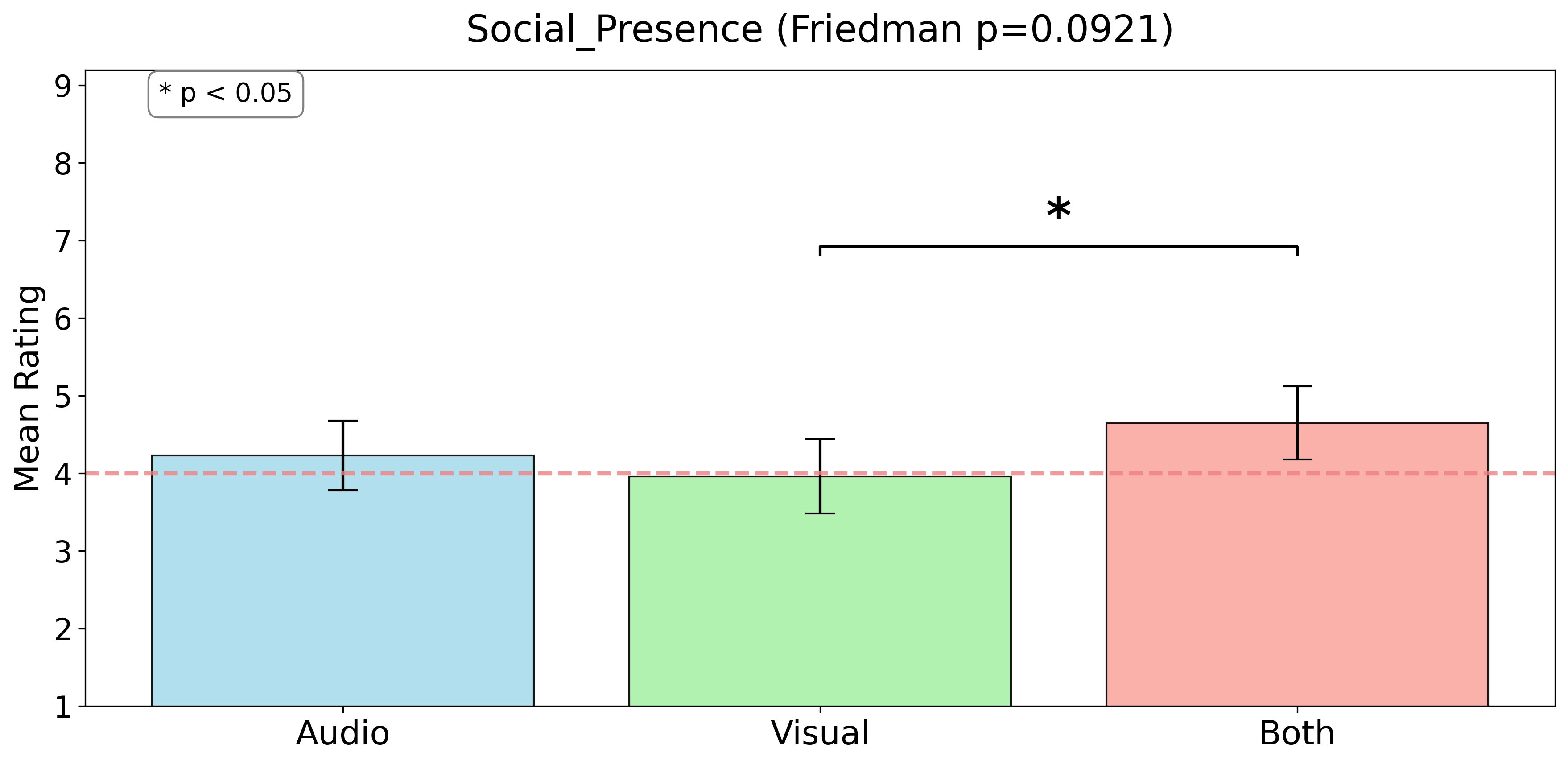}
  \caption{Pairwise testing indicates a significant difference between the Visual and Both conditions ($p = 0.015$). Red line indicates neutral score.}
  \label{fig:social_presence}
\end{figure}

The experience of Social Presence—the feeling of being connected or co-present with the other audience members—showed a marginal overall effect across the three conditions (Friedman $p = 0.092$). Post-hoc pairwise testing uncovered a significant difference specifically between the Visual and Both conditions ($p = 0.015$), driven by a large effect size ($r = 0.65$). This suggests that adding audio elements to a visual experience significantly enhances the feeling of social presence compared to experiencing visual elements alone (see Figure \ref{fig:social_presence}). 

% However, when interleaving the baseline comparisons, the data shows that despite this relative enhancement between conditions, none of the modalities independently achieved a Social Presence score that was significantly higher than the absolute neutral baseline (Audio $p = 0.637$; Visual $p = 0.898$; Both $p = 0.208$). Therefore, while the multimodal ``Both" condition is demonstrates greater social presence as compared to the ``Visual" condition, the interventions overall did not drastically exceed a baseline neutral state for this specific metric. As discussed in later sections, this finding is likely due to a combination of the voting mechanism and submission process. 

\subsubsection{Measures of Valence and Arousal Across Conditions}
% To assess the core affective impact of the conditions, participants' emotional responses were evaluated along the dimensions of Valence (the positivity of the emotion) and Arousal (the level of excitement or activation). When comparing the modalities directly against each other, no overall significant differences were observed for either Valence (Friedman $p = 0.388$) or Arousal (Friedman $p = 0.307)$. 

% Consequently, pairwise comparisons between the Audio, Visual, and Both conditions did not yield any significant relative differences for these metrics (all $p > 0.18$).

% Evaluating scores against the neutral baseline demonstrated a significant effect on emotional engagement. For Valence, both the Audio condition ($p = 0.021, r = 0.62$) and the Both condition ($p = 0.003, r = 0.80$) were rated significantly higher than the neutral baseline, exhibiting large effect sizes (see Figure \ref{fig:dav_combined}). Similarly, Arousal was significantly elevated above the baseline in the Audio condition ($p = 0.002, r = 0.84$) and the Both condition ($p = 0.004, r = 0.76$). The Visual-only condition failed to significantly exceed the neutral baseline for both Valence ($p = 0.106$) and Arousal ($p = 0.215$). These results suggest that the inclusion of audio elements is the primary catalyst for emotional engagement; without audio, visual elements alone are insufficient to elevate users' affective states—either in terms of positivity or excitement—significantly beyond a neutral experience.

Conditions containing audio successfully elevated emotional engagement above the neutral baseline. Both the Audio and ``Both" conditions significantly increased Valence (Audio: $p = 0.021, r = 0.62$; Both: $p = 0.003, r = 0.80$) and Arousal (Audio: $p = 0.002, r = 0.84$; Both: $p = 0.004, r = 0.76$) with large effect sizes (Figure \ref{fig:dav_combined}). The Visual-only condition did not significantly exceed the baseline for either metric ($p > 0.10$). However, direct comparisons between the Audio and Visual-only conditions were not statistically significant ($p > 0.32$). Thus, while audio elevates affective states above a neutral experience, it is not strictly superior to visual elements alone.

\subsubsection{Perceived Influence and Agency of Specific Audio Components}

\begin{figure}[h]
  \centering
  \includegraphics[width=\linewidth]{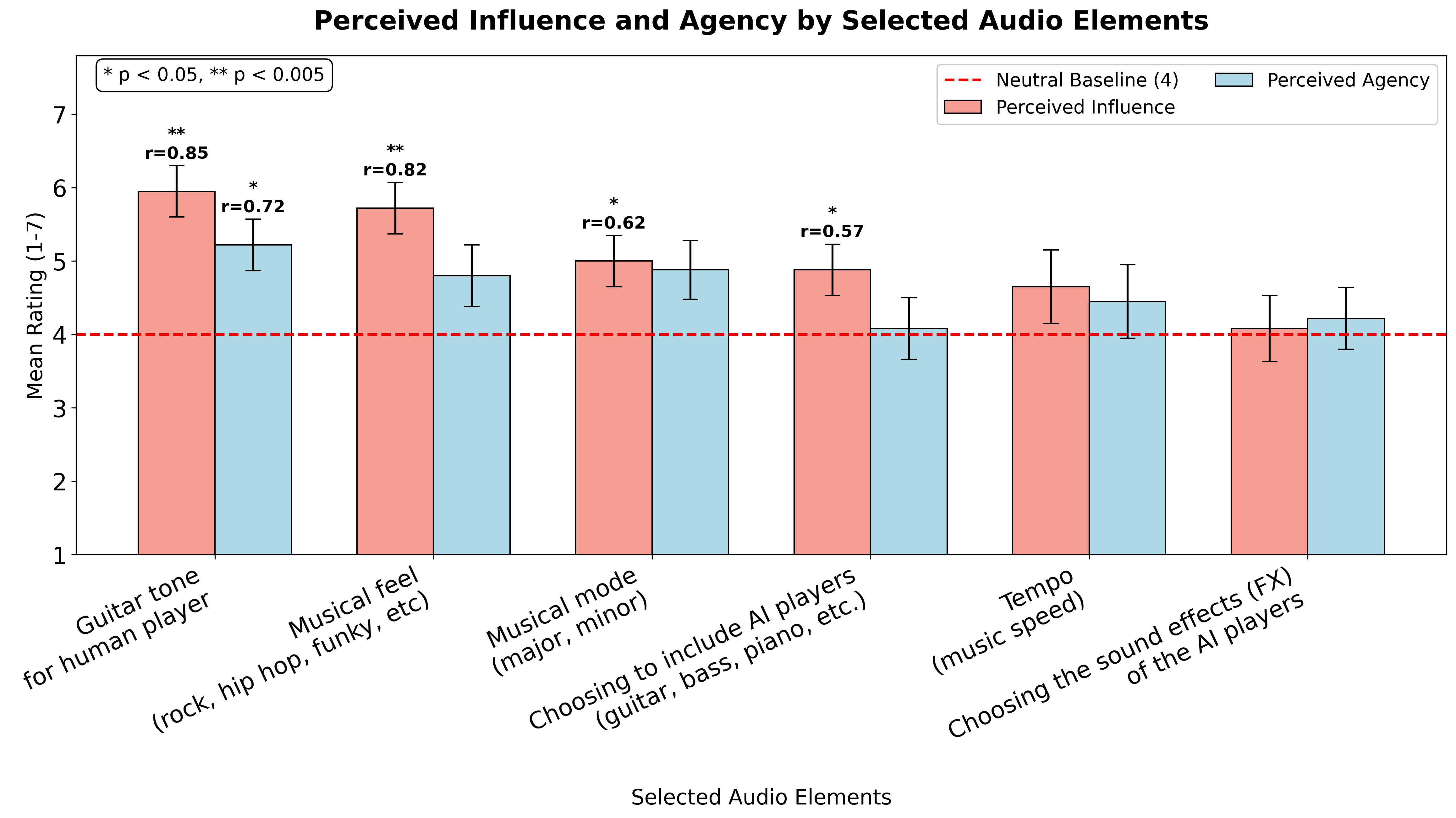}
  \caption{Indicates perceived influence and agency on the performance when selecting specific audio elements. Both measures use a rating scale of 1-7. Significance figures and effect sizes (r) are demonstrated with respect to the neutral baseline of 4 (shown in red).}
  \label{fig:audioAgencyandInfluenceVneutral}
\end{figure}

To further investigate the specific impact of sound, various audio elements were evaluated for their individual contributions to feelings of Agency and overall Influence (see Figure \ref{fig:audioAgencyandInfluenceVneutral}). When assessing which specific audio features gave participants a sense of agency, ``Guitar tone for human player" stood out as the only element rated significantly above the neutral baseline ($p = 0.007$, large effect size $r = 0.72$). Other elements, such as musical mode, tempo, and choosing AI player effects, did not significantly exceed the baseline for agency (see Figure \ref{fig:audioAgencyandInfluenceVneutral}). However, when evaluating broader influence on the experience, participants rated several core musical elements as significantly impactful compared to baseline. These included Guitar tone ($p = 0.002, r = 0.85$), Musical feel/genre ($p = 0.002, r = 0.82$), Musical mode ($p = 0.020, r = 0.62$), and Choosing to include AI players ($p = 0.032, r = 0.57$). An omnibus test confirmed an overall significant difference among the influence ratings of these specific elements (Friedman $p = 0.016$). Uncorrected pairwise p-values highlighted strong trends for core musical features over granular settings; for instance, customizing the Guitar tone was rated much more influential than choosing specific sound effects for generated instrument parts ($p = 0.009, r = 0.70$).

% or tempo ($p = 0.017, r = 0.64$).

% \begin{figure}[h]
%   \centering
%   \includegraphics[width=\linewidth]{robust_baseline_audio_agency.png}
%   \caption{Indicates perceived agency when selecting specific audio elements. Guitar tone for human player is the only rated significantly above a neutral score (4, rating scale 1-7).}
%   \label{fig:audioAgencyVneutral}
% \end{figure}

% \begin{figure}[h]
%   \centering
%   \includegraphics[width=\linewidth]{robust_baseline_audio_influence.png}
%   \caption{Indicates perceived influence on the performance when selecting specific audio elements. Guitar tone for human player, musical feel/genre, musical mode, choosing which AI instruments/band members to include rated significantly above a neutral score (4, rating scale 1-7).}
%   \label{fig:audioInfluenceVneutral}
% \end{figure}

\subsection{Thematic Analysis}
% \todo[inline]{add P#s and tally}
Thematic analysis is based on free response questions and interview data ascertained post-performance. Participants discussed agency, sense of social presence with other audience members, and commented on how to improve the performance paradigm. Both free responses and interview data were used for thematic analysis.

We recorded, transcribed the interviews, and isolated responses from the 14 participants. Transcriptions and responses were then coded using inductive thematic analysis, rooted in grounded theory \cite{badreddin2013thematic,Hopkins2023Virtual,hopkins2025mrdaw}. The first and second authors independently coded and conducted the analysis of all 14 participants. We then discussed the topics considerably and converged on the most present topics in the dataset \cite{terry2017thematic,hopkins7brain}. MAXQDA \cite{kuckartz2019analyzing_MAXQDA} software was used to keep track of independent coding databases which facilitated discussion. The coding databases were merged after discussion to produce the following final themes.

\subsubsection{Sense of Agency}
% \paragraph{Subtheme: Influence Through Voting Mechanics}
% % The voting mechanism served as the primary interface for audience members to exert their agency during the performance, leading to mixed feelings regarding personal control. For many participants, the structured nature of the voting system provided a clear, discernible way to interact with the performance, distinguishing it from traditional live music. As one participant noted in the survey, “I liked the voting aspect, it was fun seeing which options were going to win and then seeing how they impacted the show.” 

% The collective aggregation of choices often diluted the feeling of individual influence, as 
\paragraph{Subtheme: Audio Parameters Require Macro-Adjustments}
The audience's sense of agency was heavily hindered by the difficulty of tracking individual changes within the aggregated soundscape. Participants observed that while the ``big, general elements felt responsive," more granular audio settings ``weren't present enough in the music for me to tell" (P13), making it ``hard to track every parameter" (P4). This lack of clarity was compounded by the fact that aggregating disparate audio choices often led to conflicting results; for instance, combining an ``ethereal or atmospheric" vibe with incompatible instrument tones resulted in what P11 described as a ``jumbled musical idea". To reclaim a sense of control, users expressed a desire for deeper musical intervention, wishing for the ability to ``control chord progressions, drum parts and other arrangement pieces" (P3) or to dynamically ``adjust all the parameters live" (P6).

\paragraph{Subtheme: Visual Prominence and Engagement}
When analyzing the sensory feedback of the performance, the visual components often had a more immediate and noticeable impact on the audience than the audio manipulations. Participants frequently commented on the striking nature of the visual changes and how the crowd reacted to them. One participant observed, “I feel audience is more affected by the visual change, (P1)” concluding that “the visual is more controllable and noticable”. The MR components and character animations provided clarity and evidence that the audience's choices were reflected. 

\paragraph{Subtheme: Boundary Testing and Extreme Audio Selections}
A prominent sub-theme regarding agency was the audience's tendency to test the limits of the AI-mediated system rather than striving for traditional musicality. This accomplished two main aims: 1) testing the limits of the system, and 2) choosing extreme selections to maximize individual influence over the performance.  One participant indicated in their survey response, “I ended up picking extreme options to show much influence I had. (P5)” This sentiment was echoed strongly during the interviews, with one creator noting that instead of making choices that ``would make the most musical sense," the audience's mentality was often to ``take this to the max...with the scariest sound possible. (P2)" This behavior highlights a playful, almost adversarial relationship between the user and the system, where the goal was not necessarily to create a cohesive song, but to understand the boundaries of the technology and attempt to exert personal agency. The sheer novelty of the combinatorial space encouraged this experimental approach, allowing users to stretch the system until they wondered, ``when's it gonna fall apart? (P1)"

\subsubsection{Social Dynamics and Connection}

% \paragraph{Subtheme: Feelings of Social Isolation}
% % Despite the participatory nature of the event, a significant portion of the data points toward a profound lack of direct interpersonal connection among the audience members. The reliance on digital interfaces for input meant that participants were often hyper-focused on their own screens rather than engaging with the people physically around them. 

% One interviewee explicitly stated, “I definitely felt less of a connection with the people around me,” while a survey respondent similarly noted they ``felt more social collaboration than connection with others." This sentiment was corroborated in a separate interview where a participant shared, “To be honest... it was cool that we were able to vote. But it didn't particularly feel like it bonded me with other people in the audience.” The individualistic act of looking down to make selections created a somewhat siloed experience. Consequently, while the audience was co-located and participating in the same event, the technological mediation often acted as a barrier to traditional social bonding.

\paragraph{Subtheme: Emergent Collective Collaboration and Engagement}
Voting on preferences before the beginning of a set was noted as an unoptimized method of garnering social collaboration. As one participant stated, “it was cool that we were able to vote. But it didn't particularly feel like it like bonded me with other people in the audience. (P6)” Furthermore, participants suggested improvements to make this mechanic feel more dynamic and responsive, such as implementing a live voting window where users ``can see how many people are voting each option in real time (P3)" to increase active engagement. While voting successfully facilitated interaction, the resulting sense of agency was highly dependent on whether a participant's individual choice aligned with the majority and affected human performers. One survey respondent captured this sentiment, stating, “I felt we could all create a unique performance, made me more attentive. (P14)” The realization that the performer “received input from the audience and the opinion of the audience actually had an effect on the performance (P3)” fostered a sense of shared responsibility for the outcome.

\section{Themes in Musician Evaluation}
% \todo[inline]{complete musician thematic analysis and integrate into the discussion}
Musicians provided a self-reflective written statement pertaining to their experience throughout the two performances. The musicians included an improvising guitar player who played alongside the AI generated music, while the DJ was arranging and playing AI-generated clips as that were programmatically populated in Ableton. Both were practicing musicians with over 10 years of experience. The guitar player also has experience in improvised performances including technology-mediated audience participation. 

Throughout the performance there was non-verbal communication between the two musicians, particularly with regard to ending each song (roughly every 7 minutes). However, each musician made improvised musical decisions based on the generated music, set parameters chosen by the audience (e.g., tempo), and each other's respective musical choices. 

\subsection{Self-reflective Analysis of Performer Dynamics}

By examining the firsthand experiences of both the live improvising musician and the DJ, several themes emerge regarding agency, systemic constraints, and the necessity of highly adaptive collaboration. 

\subsubsection{Relinquishing Agency and Adversarial Audience Dynamics}
A primary theme characterizing the performance was the live musician's loss of creative agency to an unpredictable, audience-driven generative system. The Guitarist described the experience as stripping them of traditional control, noting, ``All the agency is taken away from me... I have almost none.'' Rather than building upon predictable elements in other improvised performances the live musician was forced to react to what they perceived as an adversarial audience actively seeking to test the performer's limits. The Guitarist observed an explicit ``push towards chaos'' from the crowd, where participants frequently selected ``intentionally wonky'' combinations simply to ``push my capabilities as far as they could possibly go.'' 

\subsubsection{Structural Mediation and Systemic Constraints}
While the Guitarist contended with the audience's extreme selections, the DJ stated that much of their attention focused on actively managing the system's output to maintain musical viability. The DJ reported feeling ``a bit limited in the ways in which I could build up the song... especially in certain genres...'' The Guitarist echoed this sentiment regarding the AI's structural homogeneity, observing that despite the parameters changing, the underlying tracks could ``feel [like they're] getting stale at times.'' These issues could be mitigated in future systems by increasing the number of musical options generated, or even enabling the audience to weigh in on ``real-time decisions to change up the music" as was mentioned by the guitar player.  

% \subsubsection{Inter-Performer Communication and Navigating Ambiguity}
% Because the generative system lacked traditional, predictable musical cues, direct on-stage communication became the critical bridge between the AI and the live improviser. The Guitarist relied heavily on visual and verbal confirmation from their co-performer, stating they were constantly ``looking at the DJ, and I'm trying to figure out okay, when are you going to launch something new?'' Recognizing this necessity, the DJ prioritized explicit verbal cues over the loud music to signal structural shifts. The DJ explained, ``The most important tracks to warn when you're going to change tend to be either the guitar, piano, or drums... I can just verbally say which track I'm changing.'' By establishing this communicative protocol, the foundational elements remained steady, which the DJ observed allowed the collaboration to go ``naturally smoothly... and good to improvise off of.''

\subsubsection{Creative Expansion and the Reward of Coherence}
Despite the distinct frustrations of navigating restricted agency, adversarial audience inputs, and system constraints, both performers reported highly rewarding creative outcomes. The sheer unpredictability of the generative outputs forced the Guitarist to combine highly disparate stylistic elements---such as blending delicate acoustic atmospheres with heavy metal distortion---which pushed them to ``creative levels... that are beyond the scope of what I would even consider doing.'' The tension between the audience's push for chaos and the performers' aim for structure resulted in many opportunities for a satisfying sense of musical cohesion. As the Guitarist reflected, ``Anytime there's a moment of coherence or a moment of, like, oh, wow, that sounds really good, that's a magical moment in an improvised jam session.'' These accounts show that while AI-mediated audience participation introduces operational friction, the necessity for adaptive collaboration yields a unique and creatively generative performance environment.

% \input{04d_DJ_Evaluation}

%%%%%%%%%%%%%%%%%%%%%%%%%%%%%%%%%%%%%%%%%%%%%%%%%%%%%%%%%%%%%%%%%%%
\section{Discussion}
% \todo[inline]{add discussion about the evaluation and limitations of the study. Add references for this section}
%\subsection{Discussion}
% \todo[inline]{maybe add a small section or edit some sections to explicitly comment on the use of AI and MR as an interactive modality rather than audio and visuals. Perhaps draw a distinction in some way.}

Our findings highlight the dynamics of multimodal audience participation in MR musical performances. A tension emerged between the affordances of generative AI, the behavioral limitations of form-based input, and the audience's desire for continuous involvement and social connection through collective decision making. By analyzing behavioral data, psychometric questionnaires, and qualitative thematic analysis, several key implications for the design of interactive live music systems became apparent.

\subsection{The Illusion of Choice: Parameter Hierarchy and Behavioral Limits}
The behavioral analysis suggests a clear cognitive \textbf{threshold regarding the time and effort participants are willing to dedicate to asynchronous, form-based input}. Despite being presented with a large combinatorial space of generative AI parameters, engagement exhibited diminishing returns. When all options were available in the ``Both'' condition, participants consistently bypassed granular AI configurations (e.g., AI instrument effects, showing skip rates up to 38.5\%) in favor of macro-structural choices like tempo (0\% skip rate) and visual characters (7.7\% skip rate) indicating a decision fatigue for less impactful parameters.

This behavioral data aligns with our subjective Influence metrics, where foundational musical elements significantly outweighed nuanced AI settings. Notably, customizing the ``Guitar tone for human player'' yielded the highest perceived influence ($p = 0.002, r = 0.85$) and was the only parameter to significantly elevate feelings of specific agency above the neutral baseline ($p = 0.007, r = 0.72$). This human-centric focus reveals that \textbf{AI in live musical contexts functions most effectively as a supplement for performance manipulation rather than the primary object of attention}. Thematic data corroborates this, indicating that audiences actively sought to challenge the live human improviser—often selecting extreme parameter combinations to test the boundaries of the performer's adaptability. Consequently, \textbf{agency was heavily dependent on observability}; participants needed to see or hear the human musician react to their specific choices to feel a genuine sense of control.

The findings suggest that perceived agency is driven less by the number of parameters offered to participants than by the \emph{observability} of their influence on the performance. Audience members consistently favored parameters whose effects were recognizable, and attributable to their actions, such as tempo, musical style, human guitar tone, or visible changes in the virtual environment. Conversely, granular AI parameters were often ignored because their individual contribution to the resulting performance was difficult to perceive. This suggests that the design of \textbf{TMAP systems should prioritize making the consequences of audience actions explicit rather than simply increasing the number of controllable parameters}.

\subsection{Sensory Asymmetries: Affective Audio vs. Cohesive Visuals}
Our results suggest \textbf{complementary roles for the audio and visual participation modalities}. Within the evaluated \systemname{} performance, audio participation was associated with affective engagement, whereas visual participation primarily provided clearly observable feedback. These findings suggest that, in this context, the audio components became the main focus of participants' engagement---especially due to the impact on human musicians---while the visual components complemented the experience by making audience influence more readily observable. Both the Audio and the ``Both'' conditions significantly elevated Valence ($p < 0.05$, $r > 0.60$), Arousal ($p < 0.01$, $r > 0.70$), and SAM Dominance ($p < 0.05$, $r > 0.60$) above a neutral baseline. Although tangential to the primary focus, and perceptibly ``programmatic'', visuals have the opportunity to provide a sense of agency at scale as they conflict less generally than audio parameters and may have had more perceived influence if more tightly coupled to a human performers output.

Auditory manipulation, is highly susceptible to structural conflict. As highlighted in the qualitative analysis, disparate musical choices frequently resulted in chaotic, dissonant noise. This is likely due to constraints imposed by the nature of sound waves, which superimpose in unpredictable ways, unlike visuals. For example, applying musical effects to various instruments results in a combinatorial explosion that is intractable to test pre-performance (e.g., 5 instruments with 5 states including ON, OFF, and effects is $5^{5}$ or 3,125 possible combinations). While some participants enjoyed the ``sheer weirdness'' of this dissonance, it often disrupted musical cohesion if too many auditory options are available. Some auditory parameters can be non-conflicting, such as genre, mode, and tempo, though these macro-structural parameters are limited. Conversely, thematic feedback revealed that visual changes (e.g., avatars, dancers, environments) were perceived as ``most apparent'' and less confounding. Because the visual domain enables multiple non-conflicting elements to be present simultaneously, it can offset the dilution of agency that participants experienced when some auditory choices were less apparent. Thus, \textbf{a multimodal system is optimal}: it utilizes audio to drive affective arousal while leveraging visuals to provide clear, parallel feedback that validates individual choices without compromising the overall aesthetic experience.

\subsection{A Conceptual Model of Perceived Agency}

Across the quantitative and qualitative findings, perceived agency depends primarily on the observability of the consequences of audience actions. We propose a conceptual model in which interactive modalities first determine the observability of system responses, which in turn shape perceived agency and ultimately influence engagement and collaboration. Furthermore, parameters whose effects are high-impact and easily attributable---such as genre, guitar tone, or visible changes in the virtual environment---produce a stronger sense of agency than parameters whose influence is subtle or distributed across multiple components. Human performers further reinforce this process because their visible adaptation provides a socially meaningful confirmation that audience input has influenced the performance. This outlook would explain why audience members were keen to test parameter boundaries especially with respect to parameters that directly impact human performers. Thus, audio and visual modalities contribute differently to the audience experience and suggests that \textbf{future TMAP systems should consider prioritizing multi-modality and observable, high-impact parameters to enhance agency}.  

% This may also partly explain why audience members were keen to test parameter boundaries, raising further questions regarding the relationship between observability and agency.

\subsection{Mediated Participation and Social Isolation}
Despite the system's collaborative premise, the asynchronous voting mechanism proved unsubstantial for fostering true social presence. Both our statistical and qualitative findings highlight that technology-mediated participation can concurrently enable mass interaction while inhibiting interpersonal connection. While the ``Both'' condition significantly improved Social Presence compared to the Visual condition ($p = 0.015, r = 0.65$), none of the three conditions managed to elevate Social Presence significantly above an absolute neutral baseline (all $p > 0.20$). The aggregation and averaging of votes obscured individual contributions, leading to frustration and a sense of isolation when participants' choices were not explicitly reflected. 

% While a novel sense of collective collaboration was observed during the performance, the interface lacked the transparency required to make users feel like they were part of a unified, socially present collective.

% This stagnation in social bonding is explained by the thematic sub-theme of ``Feelings of Social Isolation.'' The reliance on digital forms and individual voting created an atmosphere that promoted the sense of isolation. Rather than connecting with adjacent audience members, participants tended to complete forms silently and individually. Furthermore, the aggregation and averaging of votes obscured individual contributions, leading to frustration and a sense of isolation when participants' choices were not explicitly reflected. While a novel sense of collective collaboration was observed during the performance, the interface fundamentally lacked the transparency required to make users feel like they were part of a unified, socially present collective.

\subsection{Towards Continuous Co-Creation and World Building}
These insights advocate for interactive performance design to move away from episodic parameter selection toward continuous, collaborative world-building. Participants explicitly expressed an \textbf{intense desire for real-time influence and transparency of live audience voting}, suggesting that form-based voting should be replaced or augmented by dynamic, continuous interaction methods. Thematic feedback strongly supported interfaces where audiences can observe near-real-time voting trends. By integrating live, dynamic, collaborative, and transparent contributions---focusing on high-impact elements---future TMAPs can greatly enhance shared impact.

\subsection{The Role of AI and MR: Enhancing Agency and Creativity}

% The integration of generative AI provides a complex, flexible, and rich backdrop for the live musicians. Crucial to the success of this system is its underlying tokenization—the ability to systematically represent and manipulate abstract musical concepts such as genre, mode, tempo, and specific instrument tones as discrete data points. This tokenized structure allows the AI to translate asynchronous audience votes into a highly dynamic, combinatorial musical landscape. By doing so, it forces the human improviser to react to unpredictable stylistic fusions, thereby elevating the performance beyond a standard backing track into a responsive, co-created ecosystem.

The primary benefit of AI-mediated audience participation is derived from the substantial combinatorial possibility space for which audiences can leverage to participate in novel music creation while challenging the creative boundaries of human musicians. This process was described by the guitarist as seemingly adversarial, while the DJ struggled to manage the system's output to keep it viable. This tension furthered a sense of agency and arousal among the crowd, while supporting creative expansion beyond the musicians habitual range. 

% This enables audiences to push human performers beyond their habitual creative boundaries and simultaneously drive emotional arousal among the audience. 

While the auditory domain appears the primary driver of affective engagement, visual elements play an equally important role in validating audience participation. Because visual changes (such as avatars, environments, and dancer animations) can coexist without the same level of perceptual clashing as audio, they provide recognizable feedback. MR greatly enhances this possibility space by introducing spatial, non-conflicting visual layers that can be continuously altered without affecting the structural integrity of the live music. 

% To resolve the tension between collective audio averaging and the desire for individual influence, future iterations of this system should leverage the individualized capabilities of MR. Allowing audience members to bring their own headsets, or providing them with greater personal agency to navigate and customize their localized visual scene, could profoundly improve the overall sense of agency. In such a setup, even if the auditory environment is collectively averaged, users could explore the visual manifestation of the performance individually. Developing and testing this hybrid MR option—where a collectively driven audio backdrop is paired with an individualized, exploratory visual scene—represents a vital next step in understanding how to maximize both personal agency and social presence in technologically mediated live art.

\section{Limitations and Future Work}

While this study provides insights into the dynamics of audience agency and social presence in AI-mediated performances, several limitations present clear avenues for future research. Primarily, the relatively small sample size ($N=17$ for quantitative behavioral metrics), uneven audience distribution and diversity between sets, and limited number of performances necessitates larger participant pools to enhance statistical power and alleviate ordering effects between audio and visual parameters. This aspect is additionally confounded by a lack of influence of visuals on human performers presenting a further limitation. Given the focus of audio on human performer reactions and influence of human reaction on the perceived affect of the audience, future research could include improvised human dance to make the modalities more comparable. Additionally, observational data suggests a mathematical relationship between the number of configurable parameters and the total number of audience members. Future research should formally model this parameter-to-audience ratio to determine the optimal ratio required to maintain a robust sense of individual agency within crowds of varying sizes. Such studies should parse the impact of observability and agency with respect to a central versus peripheral focus.

% We also suspect performance focus plays a central role in perceived agency. Further studies should parse the impact of observability and agency with respect to a central versus peripheral focus.  

% Furthermore, the implementation of the voting interface warrants deeper investigation, particularly regarding the illusion of choice. For example, despite tempo being presented as a continuous slider encompassing a vast musical range, participants tended to elect for extreme values and reported low perceived influence. We hypothesize that providing a constrained set of discrete choices (e.g., three distinct tempo categories) rather than an infinite slider may yield a higher sense of agency by making the causal link between user input and the generated output more explicitly recognizable.

The current evaluation was constrained to a specific co-located setup, limiting our understanding of how alternate viewing modalities affect the audience experience. Future work must explore fully immersive virtual reality, individualized MR, and remote participation to test the boundaries of this paradigm. Additionally, extending this system to remote participants would be a logical next step. This would enable us to explore how physically distributed audiences experience social presence and interpersonal connection,  informing whether this technology can facilitate communal world-building outside of traditional, co-located concert environments.

\section{Conclusions}

This study evaluated a novel TMAP paradigm leveraging generative AI and MR to facilitate bidirectional interaction in live music. We assert that in TMAPs, observability is the primary driver of audience agency and engagement. Within our evaluated paradigm, MR visuals fostered agency by making audience contributions visible. Generative AI parameters, however, provided observable, macro-structural control over the musical environment, influencing the live musicians. Because audiences actively seek observable outcomes, they often used these AI parameters to test musical boundaries, introducing a compelling adversarial tension between the audience and the performers. Rather than disrupting the performance, this dynamic pushes creative limits and drives deeper audience engagement. It also generates moments of musical coherence when the musicians successfully navigate and resolve the challenges presented by the audience. Future TMAP systems should leverage observability, continuous interaction, and human performer-affecting parameters to build agency while embracing the tension-resolution dynamic to elevate collaborative live music experiences.

%\begin{table}[htbp]
%\caption{Table Type Styles}
%\begin{center}
%\begin{tabular}{|c|c|c|c|}
%\hline
%\textbf{Table}&\multicolumn{3}{|c|}{\textbf{Table Column Head}} \\
%\cline{2-4} 
%\textbf{Head} & \textbf{\textit{Table column subhead}}& \textbf{\textit{Subhead}}& \textbf{\textit{Subhead}} \\
%\hline
%copy& More table copy$^{\mathrm{a}}$& &  \\
%\hline
%\multicolumn{4}{l}{$^{\mathrm{a}}$Sample of a Table footnote.}
%\end{tabular}
%\label{tab1}
%\end{center}
%\end{table}

% \section*{Acknowledgment}

% We acknowledge the support of the MUSMET project funded by the EIC Pathfinder Open scheme of the European Commission (grant agreement n. 101184379). Views and opinions expressed are however those of the author(s) only and do not necessarily reflect those of the European Union or the European Innovation Council. Neither the European Union nor the European Innovation Council can be held responsible for them. 

%\todo[inline]{Here add any other funding agency you may have}

\bibliographystyle{IEEEtran}
\bibliography{musical_XR.bib,Turchet.bib,NMP.bib,IoMusT.bib,article_specific.bib, other.bib}

\end{document}